\pdfoutput=1
\documentclass[%twocolumn,%showpacs,preprintnumbers,
amsmath,%amssymb,aps,
prd,nofootinbib,floatfix,11pt,%preprint,
]{revtex4}

\usepackage{placeins}
\usepackage{afterpage}

\usepackage{graphicx}% Include figure filesf
\usepackage{setspace}
\usepackage{bm}% bold math
\usepackage{xcolor}

\newcommand\beq{\begin{eqnarray}}
\newcommand\eeq{\end{eqnarray}}

\def\Yuk{Y}
\def\totSUP{\Phi}
\def\backSUP{\phi}
\def\propSUP{A}

\def\lsim{\mathrel{\rlap{\lower4pt\hbox{$\sim$}}
    \raise1pt\hbox{$<$}}}                % less than or approx. symbol
\def\gsim{\mathrel{\rlap{\lower4pt\hbox{$\sim$}}
    \raise1pt\hbox{$>$}}}            

\allowdisplaybreaks
\newcommand\lnbar{\overline{\ln}}
\newcommand\MSbar{$\overline{\rm{MS}}$ }

\newcommand{\sdfrac}[2]{\mbox{\small$\displaystyle\frac{#1}{#2}$}}

\begin{document}

\renewcommand{\theequation}{\arabic{section}.\arabic{equation}}
\renewcommand{\thefigure}{\arabic{section}.\arabic{figure}}
\renewcommand{\thetable}{\arabic{section}.\arabic{table}}

\title{\Large\baselineskip=20pt 
Effective K\"ahler and auxiliary field potentials\\ for chiral superfield models at three loops 
}
\author{Stephen P.~Martin}
\affiliation{\mbox{\it Department of Physics, Northern Illinois University, DeKalb IL 60115}}

\begin{abstract}\normalsize \baselineskip=15.5pt 
I obtain the effective K\"ahler potential at three-loop order for a general renormalizable supersymmetric theory containing only chiral supermultiplets. The three-loop contribution is remarkably simple, consisting of only four terms involving three distinct renormalized master integrals. In the case of the Wess-Zumino model with a single chiral superfield, I also obtain the effective auxiliary field potential at three-loop order, extending previous results at one-loop order. The method used is inferential, relying on existing knowledge of the ordinary scalar effective potential.  
\end{abstract}

\maketitle

\tableofcontents

\baselineskip=14.6pt

\newpage

%%%%%%%%%%%%%%%%%%%%%%%%%%%%%%%%%%%%%%%%%%%%%%%%%%%%%%%%%%%%%%%
\section{Introduction\label{sec:intro}}
\setcounter{equation}{0}
\setcounter{figure}{0}
\setcounter{table}{0} 
\setcounter{footnote}{1}

Radiative corrections in quantum field theories can be summarized in terms of the effective action, which
can be computed in perturbation theory by summing 1-particle irreducible vacuum Feynman diagrams in the presence of external background fields. For some purposes, it is sufficient to have the scalar effective potential \cite{Coleman:1973jx,Jackiw:1974cv,Sher:1988mj}, which is the effective action in the case that the background fields are taken to be independent of spacetime position. At present, the effective potential is known for a general renormalizable theory through 2-loop order \cite{Ford:1992pn,Martin:2001vx} and 3-loop order
\cite{Martin:2013gka,Martin:2017lqn}, including for supersymmetric gauge theories \cite{Martin:2023fno}, and with leading QCD corrections for the Standard Model at 4-loop order \cite{Martin:2015eia}.

Despite increasingly strong constraints from collider  and dark matter experiments, supersymmetry remains
a viable possibility for physics beyond the Standard Model, and it arises in string theories that propose to
consistently quantize gravity. It is therefore useful to understand as much
as possible about the form of radiative corrections in the supersymmetric context. In particular, the Higgs vacuum expectation values
in supersymmetric extensions of the Standard Model are most easily determined from the effective potential, and high-order radiative corrections are clearly necessary in this regard. The radiative corrections to the effective Lagrangian for supersymmetric theories have a more constrained structure than in ordinary generic quantum field theories, and one motivation for the present work is to take a step to learn more about this, although supersymmetric gauge theories are not treated in the present paper. Indeed, it is found below that the effective K\"ahler potential at three-loop order does not include one term that might otherwise be expected to be present.

For supersymmetric models, the effective action can be written as an integral over superspace of background superfields. In this paper, I will concentrate on renormalizable models that contain only chiral superfields. I will also consider only the part of the superfield effective action that contains no spacetime derivatives, so that the results imply the physical content of the scalar effective potential. In terms of background chiral superfields $\phi_i$ satisfying $\partial_\mu \phi_i = 0$,
the effective Lagrangian can be written in the form\footnote{For reviews of supersymmetry using the notations and conventions to be followed in this paper, see refs.~\cite{Martin:1997ns,Dreiner:2023yus}.}
\beq
{\cal L}_{\rm eff} &=& \int d^2\theta d^2\theta^\dagger\, 
L_{\rm eff}(\phi_i,\, \phi^{*i},\, D_\alpha \phi_i,\, \overline D^{\dot\alpha}\! \phi^{*i},\,
DD\phi_i,\, \overline D\overline D \phi^{*i}) + 
\left ( \int d^2 \theta\, W(\phi_i) + {\rm c.c.} \right )
,
\eeq 
where $W(\phi_i)$ is the superpotential, and it is conventional to split $L_{\rm eff}$ into two parts,
\beq
L_{\rm eff} = K_{\rm eff} + J_{\rm eff}.
\eeq
Here $K_{\rm eff}(\phi_i,\, \phi^{*i})$ is called the effective K\"ahler potential,
with a canonical tree-level part $\phi^{*i} \phi_i$, and is defined to be the part that contains no chiral superderivatives $D_\alpha$ or $\overline{D}^{\dot\alpha}$. The remaining part $J_{\rm eff}$ is thus defined by the property that it vanishes when all chiral superderivatives of $\phi_i$ and $\phi^{*i}$ are replaced by 0. It is often called the effective auxiliary field potential \cite{Kuzenko:1998tsq} because, when evaluated in terms of constant background bosonic component fields, it contains at least three $F$-term auxiliary fields. The effective superpotential $W(\phi_i)$ is the same as the tree-level one.\footnote{In this paper, I assume that all of the chiral superfields are treated as massive, so that nonrenormalization theorems \cite{Grisaru:1979wc} apply, forbidding all (even finite) perturbative quantum corrections to the superpotential. For exactly massless chiral superfields, the proof of the nonrenormalization theorem has a loophole, and there can be nondivergent perturbative corrections to the effective superpotential starting at 2-loop order, as argued in ref.~\cite{West:1990rm} and shown explicitly in refs.~\cite{Jack:1990pd,Buchbinder:1994xq}.}

The effective K\"ahler potential and the effective auxiliary field potential can be calculated as loop integrals either in terms of component fields or using superspace methods. The present state of the art
is that the effective K\"ahler potential is known at 1-loop \cite{Buchbinder:1994iw,deWit:1996kc,Pickering:1996he,Grisaru:1996ve,Brignole:2000kg,GrootNibbelink:2005nez} and 2-loop \cite{GrootNibbelink:2005nez}
orders for general supersymmetric gauge theories including non-renormalizable ones, while the effective auxiliary field potential is known \cite{Kuzenko:2014ypa,Tyler:2013mgu}
at 1-loop order for the Wess-Zumino model with a single chiral superfield.
In this paper, I will extend these results by computing the 3-loop effective K\"ahler potential for a renormalizable theory with an arbitrary number of interacting chiral superfields, and the 3-loop effective auxiliary field potential for the Wess-Zumino model with a single chiral superfield. Instead of calculating
these superfield effective potentials directly, I will infer them by leveraging the existing knowledge of the 3-loop ordinary scalar effective potential. As explained in more detail below, this is done by expanding the scalar effective potential in terms of the supersymmetry breaking auxiliary fields. The quadratic part of this expansion in auxiliary fields gives the effective K\"ahler potential, while the whole scalar effective potential is needed to find the effective auxiliary field potential. In practical applications, the effective K\"ahler potential is often sufficient when loop corrections are important but spontaneous supersymmetry breaking can be treated as a subdominant effect.

\begin{figure}[t]
\begin{center}
\includegraphics[width=0.93\linewidth,angle=0]{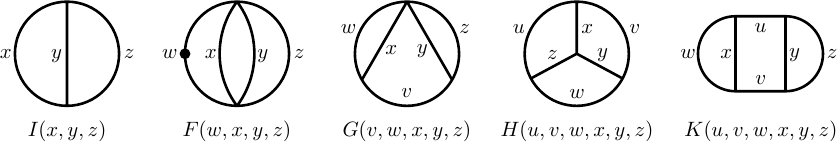}
\end{center}
\vspace{-0.35cm}
\begin{minipage}[]{0.96\linewidth}
\caption{\label{fig:topologies}
Feynman diagram topologies corresponding to the renormalized $\epsilon$-independent scalar vacuum integrals used to express the results in this paper, with squared-mass labels $u,v,w,x,y,z$. Reference \cite{Martin:2016bgz} provides the specific definitions of the integral functions, the differential equations that they satisfy, known analytical results in special cases, and a computer program {\tt 3VIL} allowing for their fast and accurate numerical evaluation in general.}
\end{minipage}
\end{figure}

The results below are expressed in the modified minimal subtraction ($\overline{\rm{MS}}$) renormalization scheme based on dimensional regularization
in $d = 4 - 2 \epsilon$ dimensions. (Since no gauge interactions are involved, it is not necessary to make a distinction between dimensional regularization and dimensional reduction.) One-loop order radiative corrections result in logarithms, which will be written as
\beq
\lnbar(x) &\equiv& \ln(x/Q^2),
\eeq
where $x$ is the squared mass in the loop integration and $Q$ is the \MSbar renormalization scale,
which is related to the loop momentum dimensional regularization mass scale $\mu$ by
\beq
Q^2 &=& 4\pi e^{-\gamma_E} \mu^2.
\eeq  
At 2-loop order, I make use of the renormalized $\epsilon$-independent master vacuum integral $I(x,y,z)$,
as defined for example in eq.~(5.4) of ref.~\cite{Martin:2016bgz} in terms of logarithms and dilogarithms. At three-loop order, the results below similarly depend on renormalized $\epsilon$-independent master vacuum integrals 
$F(w,x,y,z)$, $G(v,w,x,y,z)$, and $H(u,v,w,x,y,z)$, which cannot (in general) be expressed analytically in terms of classical polylogarithms. It is also convenient to define the integral combinations
\beq
\overline I (w,x,y,z) &=& 
\left [I(w,y,z) - I(x,y,z) \right ]/(x-w) \qquad \mbox{(for $x \not= w)$},
\\
\overline I (x,x,y,z) &=& -\frac{\partial}{\partial x} I(x,y,z),
\\
K(u,v,w,x,y,z) &=& \left [G(u,w,x,y,z) - G(v,w,x,y,z) \right ]/(v-u) \qquad \mbox{(for $u \not= v)$},
\\
K(v,v,w,x,y,z) &=& -\frac{\partial}{\partial v} G(v,w,x,y,z).
\eeq
All of these integrals functions have an explicit logarithmic dependence on the renormalization scale $Q$, although it is not listed explicitly among
the arguments because it is always the same everywhere within a given calculation. 
The corresponding Feynman diagram topologies are shown in Figure \ref{fig:topologies},
%%%%%%%%%%%%%%%%%%%%%%%%%%%%%%%%%%%%%%%%%%%%%%%
%%%%%%%%%%%%%%%%%%%%%%%%%%%%%%%%%%%%%%%%%%%%%%%%%%%%%%%%%%%%%%%%%%%%
and the precise definitions are found in ref.~\cite{Martin:2016bgz}, which also provides the differential equations that they satisfy, and a computer program {\tt 3VIL} allowing for their fast and accurate numerical evaluation. These same master integrals were used to express the closely related results for the 3-loop effective potential in a general renormalizable theory in ref.~\cite{Martin:2017lqn}.
In the special case that the squared mass arguments are all equal, $I(x,x,x)$, $\overline{I}(x,x,x,x)$, $F(x,x,x,x)$, $G(x,x,x,x,x)$, $H(x,x,x,x,x,x)$, and $K(x,x,x,x,x,x)$ can each be expressed analytically in terms of $\lnbar(x)$ and transcendental constants, as given in ref.~\cite{Martin:2016bgz} and in the next section of the present paper.

\section{Effective K\"ahler potential at three loops\label{sec:Kahler}}
\setcounter{equation}{0}
\setcounter{figure}{0}
\setcounter{table}{0} 
\setcounter{footnote}{1}

Consider a supersymmetric model with chiral superfields $\totSUP_i$ ($i = 1,2,\ldots,N_s$) with a superpotential
\beq
W(\totSUP) &=& \frac{1}{2} \mu^{ij} \totSUP_i \totSUP_j + \frac{1}{6} \lambda^{ijk} \totSUP_i \totSUP_j \totSUP_k ,
\label{eq:WXi}
\eeq 
and a canonical tree-level K\"ahler potential, $K = \totSUP^{*i} \totSUP_i$. [Repeated indices in this paper are implicitly summed over except when they appear on both sides of an equality. Also, flipping the index heights
on a parameter will denote complex conjugation, so that for example $\mu_{ij} \equiv (\mu^{ij})^*$.]
Now divide each chiral superfield into a background chiral superfield $\backSUP_i$ and a propagating quantum chiral superfield part $\propSUP_i$, as
\beq
\totSUP_i &=& \backSUP_i + \propSUP_i.
\label{eq:XiphiPhi}
\eeq
To define the effective K\"ahler potential, the background chiral superfields are taken to obey their classical equations of motion, 
\beq
\frac{1}{4} \overline{D}\overline{D} \backSUP^{*i} 
&=&  
\frac{\partial W}{\partial \phi_i} 
\>=\>
\mu^{ij} \backSUP_{j} + \frac{1}{2} \lambda^{ijk} \backSUP_{j} \backSUP_{k}
\>\equiv\>
w^i 
.
\label{eq:superfieldconstraint}
\eeq
Now one can rewrite the mixed terms in the Lagrangian coming from the K\"ahler potential using
\beq
\int d^2\theta d^2\theta^\dagger \phi^{*i} A_i &=& \int d^2 \theta\, \left ( -\sdfrac{1}{4} \overline D \overline D \phi^{*i} \right ) A_i,
\eeq
which, applying eq.~(\ref{eq:superfieldconstraint}), cancels the contribution linear in $A_i$ from the superpotential. Thus the propagating chiral superfields have a canonical tree-level K\"ahler potential $A^{*i} A_i$
and superpotential
\beq
W(\propSUP) &=& \frac{1}{2} m^{ij} \propSUP_i \propSUP_j + \frac{1}{6} \lambda^{ijk} \propSUP_i \propSUP_j \propSUP_k ,
\eeq
where the background-chiral-superfield-dependent mass matrix is
\beq
m^{ij} &=& \mu^{ij} + \lambda^{ijk} \backSUP_k.
\label{eq:mij}
\eeq
For the purposes of calculating or expressing the results of loop corrections in the general case with $N_s>1$, it is convenient to rotate to a new basis such that the propagating superfield squared-mass matrix is diagonal. Thus, one defines new chiral superfields $\tilde \propSUP_j$ according to
\beq
\propSUP_i &=& U_i{}^j \tilde\propSUP_j,
\label{eq:defineUA}
\eeq
where $U_i{}^j$ is a unitary matrix chosen so that
\beq
(U^\dagger)_i{}^k m_{kn} m^{np} U_p{}^j &=& \delta_i^j x_i,
\eeq
where $(U^\dagger)_i{}^k = (U_k{}^i)^*$. This is always possible because $m_{kn} m^{np}$ is a Hermitian
matrix, with eigenvalues that we are denoting by $x_i$. One can then write the superpotential for the propagating superfields as
\beq
W(\propSUP) &=& \frac{1}{2} M^{ij} \tilde\propSUP_i \tilde\propSUP_j + \frac{1}{6} Y^{ijk} \tilde\propSUP_i \tilde\propSUP_j \tilde\propSUP_k ,
\eeq
where
\beq
M^{ij} &=& m^{kl} U_k{}^i U_l{}^j,
\\
Y^{ijk} &=& \lambda^{npq} U_n{}^i U_p{}^j U_q{}^k.
\label{eq:defineUY}
\eeq
Note that the tree-level K\"ahler potential for $\tilde A_i$ is still canonical, due to the unitarity of $U$.
In general (for $N_s>1$),  $U$ depends on the background chiral superfields in a non-linear way, so the field-dependent parameters $M^{ij}$ and $Y^{ijk}$ do also, even though the latter is dimensionless. The superpotential
does not suffer radiative corrections as long as the fields are not massless, but radiative corrections to the effective K\"ahler potential are non-trivial and depend on the background chiral superfields only through the combinations $M^{ij}$ and $Y^{ijk}$ and their complex conjugates $M_{ij}$ and $Y_{ijk}$. Note that while $M_{ik} M^{kj} = \delta_i^j x_i$ is diagonal (by construction, through the choice of $U$), 
and $M^{ij}$ can also always be made diagonal, it is sometimes convenient to choose $M^{ij}$ non-diagonal if there are degeneracies. For example, this can occur if two chiral superfields
carry opposite conserved charges that are left unbroken by the background fields.

The effective K\"ahler potential and the superpotential for the background chiral superfields encode the information about the part of the ordinary effective potential $V_{\rm eff}$ obtained by expanding through second order in the supersymmetry-breaking auxiliary fields. To see this, write
\beq
{\cal L}_{\rm eff} &=&  
\int d^2\theta d^2\theta^\dagger \, K_{\rm eff}(\backSUP_i, \backSUP^{*i})
+ 
\Bigl (\int d^2\theta \> W(\backSUP_i) + {\rm c.c.} \Bigr )
.
\label{eq:Keffsuperspace}
\eeq
After integrating out the auxiliary fields, and dropping terms with fermions and space-time derivatives, one obtains
\beq
V_{\rm eff} &=& -{\cal L}_{\rm eff} \>=\> 
w^{i} \left [\biggl (\frac{\partial^2 K_{\rm eff}}{\partial \backSUP^*\partial\backSUP}  \biggr )^{-1} \right ]_i^{\phantom{..}j}
w^*_j + {\cal O}(w^3),
\label{eq:VEff}
\eeq
where the matrix ${\partial^2 K_{\rm eff}}/{\partial \backSUP^{*i}\partial\backSUP_j}$ is the effective K\"ahler metric, and $w^i$ is defined by eq.~(\ref{eq:superfieldconstraint}), and it is understood that all $\backSUP^{*i}$ and $\backSUP_i$ are replaced by their scalar components.
Terms with more than two of $w^i$ or $w_j^*$ in the scalar effective potential $V_{\rm eff}$ arise when the effective K\"ahler potential term in eq.~(\ref{eq:Keffsuperspace}) is augmented to include terms with chiral derivatives $D$ and $\overline{D}$; these terms constitute the effective auxiliary field potential, and will be discussed in the next section.

Consider a loop expansion for the effective K\"ahler potential,
\beq
K_{\rm eff} &=& \backSUP^{* i} \backSUP_i  + \sum_{\ell=1}^\infty \kappa^\ell K^{(\ell)}
,
\label{eq:Keffloopexp}
\eeq
where
\beq
\kappa = {1}/{16 \pi^2}
\eeq
will serve as the loop-counting parameter in the following. Then the inverse K\"ahler metric can be constructed iteratively in $\ell$, with the result that eq.~(\ref{eq:VEff}) can be written as
\beq
V_{\rm eff} &=& w^{i}w^*_i + \sum_{\ell = 1}^\infty \kappa^\ell w^{i} V_i^{(\ell) j} w^*_j
+ {\cal O}(w^3), 
\label{eq:Veffff}
\eeq
where
\beq
V^{(1)j}_i &=& -K_i^{(1)j},
\label{eq:V1ij}
\\
V^{(2)j}_i &=& -K_i^{(2)j} + K_i^{(1)k} K_k^{(1)j},
\\
V^{(3)j}_i &=& -K_i^{(3)j} + K_i^{(2)k} K_k^{(1)j} + K_i^{(1)k} K_k^{(2)j}
- K_i^{(1)k} K_k^{(1)n} K_n^{(1)j} ,
\label{eq:V3ij}
\eeq
etc., with
\beq
K_i^{(\ell)j} &=& \frac{\partial^2 K^{(\ell)}}{\partial \backSUP^{*i} \partial \backSUP_j} ,
\eeq
in which it is again understood that the background chiral superfields are replaced by their scalar components.
Thus, by evaluating the quadratic part of the scalar effective potential in an expansion in $w^{i}$ and $w_j^*$, one can infer the functional form of the corresponding effective K\"ahler potential contributions from their derivatives. 
Note that this is really equivalent to an expansion in supersymmetry breaking auxiliary fields, which are proportional to $w^i$ and $w_j^*$. The key point is that the functions of complex scalars in eqs.~(\ref{eq:V1ij})-(\ref{eq:V3ij}) can be ``lifted" to superspace by replacing them by the corresponding chiral superfields. 

It remains to evaluate the scalar effective potential $V_{\rm eff}$, which I do using a component-field calculation, taking advantage of the fact that for a general renormalizable theory the results at 3-loop order for
\beq
V_{\rm eff} &=& \sum_{\ell=0}^\infty \kappa^\ell V^{(\ell)},
\eeq
have been given in ref.~\cite{Martin:2017lqn} in the \MSbar scheme. This allows evaluation of eqs.~(\ref{eq:V1ij})-(\ref{eq:V3ij}) by expanding $V_{\rm eff}$ in $w^i$ and $w_j^*$. To apply the general results to the present case, 
one starts with the same superpotential of eq.~(\ref{eq:WXi}), and implements the same shift in eq.~(\ref{eq:XiphiPhi}), but with the important difference that now 
$\backSUP_i$ is taken to be the background scalar component only, satisfying $DD\phi = \overline D \overline D \phi^* = 0$, rather than a chiral superfield satisfying eq.~(\ref{eq:superfieldconstraint}). The resulting Lagrangian for the propagating fields in the presence of the scalar background fields is described by the superpotential
\beq
W(A) &=& w^i A_i + \frac{1}{2} m^{ij} A_i A_j + \frac{1}{6} \lambda^{ijk} A_i A_j A_k,
\label{eq:WAifi}
\eeq
from which one readily obtains the propagating component-field masses and interactions. Here $w^i$ and $m^{ij}$
have exactly the same form as given in eqs.~(\ref{eq:superfieldconstraint}) and (\ref{eq:mij}), but now with $\phi_i$ taken to be
complex scalar background fields. By dimensional analysis, it is clear that $w^i$ and its complex conjugate
$w_i^*$ have no effect on the field-dependent 
fermion masses and Yukawa, scalar cubic, and scalar quartic couplings, and enter only in the squared-mass terms in the tree-level potential for the complex scalar components $a_i$ of the propagating superfields $A_i$. These squared-mass terms have the form
\beq 
\frac{1}{2} 
\begin{pmatrix} a^{*i} & a_i \end{pmatrix}
\begin{pmatrix} 
m_{ik} m^{kj} & \phantom{x}\lambda_{ijk} w^k \\
\lambda^{ijk} w_k^* & \phantom{x} m^{ik} m_{kj}
\end{pmatrix}
\begin{pmatrix} a_j \\  a^{*j} \end{pmatrix}
.
\label{eq:M2s}
\eeq
The diagonal blocks in this scalar squared-mass matrix are the same as the propagating fermion squared mass
matrices, and the off-diagonal blocks involving $w^k$ and $w_k^*$ can be treated in a perturbative expansion of the general results, using the known derivatives of the master integrals. If one sets $w^k = w_k^* = 0$, corresponding to the leading term in this expansion, then supersymmetry is unbroken and the effective potential vanishes at each loop order \cite{Zumino:1974bg}. Continuing in the expansion of $V_{\rm eff}$,
terms linear in $w^k$ or $w_k^*$ and quadratic terms proportional to $w^k w^n$ and $w_k^* w_n^*$ must also vanish, as can be seen from eq.~(\ref{eq:Veffff}). This provided a check of the calculations leading to the results about to be described.

I used the general form for the effective potential of the theory described by eq.~(\ref{eq:WAifi}) in the special cases $N_s=1$ and $N_s=2$, as this allowed analytic diagonalization of the superpotential masses and is sufficient to uniquely fix the coefficients of candidate terms
in the effective K\"ahler potential. However, the results should be valid for any $N_s$. Expanding to quadratic order in $w^k$ and $w_k^*$, I infer that the effective K\"ahler potential contributions up to 3-loop order are 
\beq
K^{(0)} &=& \phi^{*i} \phi_i
,
\label{eq:K0gen}
\\
K^{(1)} &=& \sum_i x_i \Bigl [1 - \frac{1}{2} \lnbar(x_i) \Bigr ],
\label{eq:K1gen}
\\
K^{(2)} &=& \frac{1}{6} \Yuk^{ijk} \Yuk_{ijk} I(x_i,x_j,x_k),
\label{eq:K2gen}
\\
K^{(3)} &=& 
-\frac{1}{8} \Yuk^{ijk} \Yuk_{iln} \Yuk^{lnp} \Yuk_{jkp} G(x_i,x_j,x_k,x_l,x_n)
\nonumber \\ &&
+\frac{1}{4} \Yuk^{ijl} \Yuk^{i'pk} \Yuk_{jnk} \Yuk_{ln'p} M_{ii'} M^{nn'} H(x_i,x_j,x_k,x_l,x_n,x_p)
\nonumber \\ && 
+ \frac{1}{8} \Yuk^{ijk} \Yuk_{ijl} \left [ \Yuk^{lnp} \Yuk_{knp}\hspace{0.6pt} x_k 
+  \Yuk_{l'np} \Yuk^{k'np} M_{kk'} M^{ll'} \right ] K(x_k,x_l,x_i,x_j,x_n,x_p) ,
\label{eq:K3gen}
\eeq
in terms of renormalized vacuum integral functions defined in ref.~\cite{Martin:2016bgz}.
The result for $K^{(1)}$ agrees with the calculations of ref.~\cite{Buchbinder:1994iw,deWit:1996kc,Pickering:1996he,Grisaru:1996ve,Brignole:2000kg,GrootNibbelink:2005nez},
and $K^{(2)}$ successfully reproduces the 2-loop result given by S.~Groot Nibbelink and T.S.~Nyawelo 
in ref.~\cite{GrootNibbelink:2005nez} using a very different method based on direct evaluation of supergraphs.
The result $K^{(3)}$ is new, and is still remarkably simple, consisting of only four terms involving three distinct 3-loop renormalized master integrals. Indeed, it seems to be simpler than naive expectation, as at present I cannot offer an explanation for the absence of an otherwise-plausible term proportional to 
$\Yuk^{ijk} \Yuk_{ijl} \Yuk^{lnp} \Yuk_{knp}\hspace{0.6pt} x_i K(x_k,x_l,x_i,x_j,x_n,x_p)$, other than the 
empirical fact that its coefficient vanishes. Presumably this could be explained in some elegant way in a direct superfield 3-loop calculation.

In the special case of the Wess-Zumino model with $N_s = 1$ and 
\beq
W = \frac{1}{2} \mu \Phi^2 + \frac{1}{6} \lambda \Phi^3, 
\label{eq:WWZ}
\eeq
the unitary matrix $U$ is simply the number 1,
and the 3-loop integrals can be evaluated analytically \cite{Avdeev:1995eu,Broadhurst:1998rz,Fleischer:1999mp,Schroder:2005va}
in terms of logarithms and transcendental constant numbers, and are given in the conventions and notations of the present paper in ref.~\cite{Martin:2016bgz}. These integral functions are\footnote{Note that by definition $K(x,x,x,x,x,x) = 
\bigl [ -\frac{\partial}{\partial y} G(y,x,x,x,x) \bigr ] \bigr |_{y=x}$, from which eq.~(\ref{eq:Kxxxxxx}) can be obtained using the derivative given for general arguments in the ancillary file  {\tt derivatives.txt} of ref.~\cite{Martin:2016bgz}. The other necessary results were given in eqs.~(5.9), (5.31), (5.51), and (5.61) of that reference.}
\beq
I(x,x,x) &=& x \left [c_I -\frac{15}{2} + 6 \lnbar(x) - \frac{3}{2}\lnbar^2(x) \right ],
\\
G(x,x,x,x,x) &=& x 
\left [4 c_I + 6 \zeta_3 - \frac{97}{3} + (26 - 2 c_I) \lnbar(x) - 8 \lnbar^2(x) + \lnbar^3(x) \right ]
,
\\
H(x,x,x,x,x,x) &=& c_H - 6 \zeta_3 \lnbar(x),
\\
K(x,x,x,x,x,x) &=& \frac{2}{3} c_I + 2 \zeta_3 - \frac{5}{3} + \frac{2}{3} c_I \lnbar(x) + \lnbar^2(x) - \frac{1}{3} \lnbar^3(x),
\label{eq:Kxxxxxx}
\eeq
where the transcendental numerical constants are
\beq
\zeta_3 &=& \sum_{n=1}^\infty 1/n^3 \>\approx\> 1.2020569031595942854
,
\\
c_I &=& 
3 \sqrt{3}\, {\rm Im} \left [ {\rm Li}_2 (e^{2 \pi i/3}) \right ] 
\>\approx\>
3.515860858034188
,
\\
c_H &=& 16 {\rm Li}_4(1/2) - \frac{17\pi^4}{90} + \frac{2}{3} \ln^2(2) [ \ln^2(2) - \pi^2] 
+ 6 \zeta_3 - \frac{c_I^2}{3}
\>\approx\>
-10.035278479768789.\phantom{xxx}
\eeq
Thus the specialization of eqs.~(\ref{eq:K0gen})-(\ref{eq:K3gen}) to the 3-loop effective K\"ahler potential of the Wess-Zumino model is
\beq
K^{(0)} &=&  |\phi|^2,
\label{eq:K0WZ}
\\
K^{(1)} &=&  x \Bigl [1 - \frac{1}{2} \lnbar(x) \Bigr ],
\label{eq:K1WZ}
\\
K^{(2)} &=& 
|\lambda|^2 x \left [\frac{1}{6} c_I - \frac{5}{4} + \lnbar(x) - \frac{1}{4} \lnbar^2(x) \right ]
,
\\[4pt]
K^{(3)} &=& 
|\lambda|^4 x \biggl [\frac{29}{8} + \frac{1}{4} c_H - \frac{1}{3} c_I  - \frac{1}{4} \zeta_3
+ \left (\frac{5}{12} c_I -\frac{13}{4} - \frac{3}{2} \zeta_3 \right ) \lnbar(x)
\nonumber \\ && 
+ \frac{5}{4} \lnbar^2(x) - \frac{5}{24} \lnbar^3(x)
\biggr ]
,
\label{eq:K3WZ}
\eeq
where $x = |\mu + \lambda \phi|^2$. The 2-loop result $K^{(2)}$ agrees with that found in ref.~\cite{GrootNibbelink:2005nez}, 
except that in eq.~(54) of the arXiv version [eq.~(5.6) of the journal version]
the term $12 \kappa(\bar x)$  with $\bar x = 4/\sqrt{3}$ is incorrect and should actually be $2 c_I/3$ in the notation of the present paper. In the arXiv version only, there is also an obvious missing factor of $|\lambda|^2$. My result for $K^{(2)}$ also agrees with that found in ref.~\cite{Tyler:2013mgu} except that the term 
$-\zeta(2)$ in eq.~(3.2.31) of that reference should be absent. 
%In both cases the discrepancies seem to be due to (different) incorrect evaluations of the non-logarithmic part of the renormalized master integral $I(x,x,x)$.

\section{Effective auxiliary field potential at three loops\label{sec:EAFP}}
\setcounter{equation}{0}
\setcounter{figure}{0}
\setcounter{table}{0} 
\setcounter{footnote}{1}

The known scalar effective potential $V_{\rm eff}$ can also be used to infer the rest of the effective superspace Lagrangian that involves superderivatives but not spacetime derivatives acting on chiral superfields, by expanding beyond second order in $w^k$ and $w^*_k$. For simplicity, I will illustrate this in the special case of the Wess-Zumino model with one chiral superfield, with the superpotential given in eq.~(\ref{eq:WWZ}). (Although there should be no fundamental obstacle to carrying out the same procedure for
several multiplets, I have not done so, as it becomes considerably more complicated.) The effective superspace Lagrangian for the chiral superfield $\phi$ (taken to be constant in spacetime but not constant in superspace) can be written as 
\beq
{\cal L}_{\rm eff} &=& \int d^2\theta d^2\theta^\dagger \, \Bigl [ K_{\rm eff}(\phi, \phi^*) + 
J_{\rm eff}(m,\, m^*,\, \Delta,\,  \Delta^*) \Bigr ]
+
\left ( \int d^2\theta\, W(\phi) + {\rm c.c.} \right ), 
\phantom{xxx}
\label{eq:KJWeffWZ}
\eeq
where $J_{\rm eff}$ is the effective auxiliary field potential. This uses the fact that,
because of the form of the propagating superfield superpotential, $J_{\rm eff}$ can only depend on 
$\phi$ and $\phi^*$ through the background chiral superfields
\beq
m &=& \mu + \lambda \phi, \qquad\quad 
\Delta^* \>=\> -\frac{1}{4} \overline{D}\overline{D} m^*
\eeq
and their antichiral superfield conjugates
\beq
m^* &=& \mu^* + \lambda^* \phi^*,\qquad\quad \Delta \>=\> -\frac{1}{4} DD m
.
\eeq
The effective K\"ahler potential $K_{\rm eff}$ is defined to include all of the non-zero contributions for vanishing chiral derivatives, so that $J_{\rm eff}$ vanishes for $\Delta = \Delta^* = 0$. As we have already seen in eqs.~(\ref{eq:K1WZ})-(\ref{eq:K3WZ}), the loop corrections to $K_{\rm eff}$ can only depend on $\phi$ and $\phi^*$ through the combination $x = |m|^2$.  Note that it is not necessary to include a dependence on $Dm Dm$ or 
$\overline D m^* \overline D m^*$ in $J_{\rm eff}$ in the case of the Wess-Zumino model, because of the identities   
\beq
m^n (Dm Dm) (DDm)^p  &=& -\frac{1}{n+1} m^{n+1} (DDm)^{p+1}  + \cdots,
\label{eq:DphiDphiIBP}
\\ 
(m^{*})^n  (\overline D m^* \overline D m^*) (\overline D \overline D m^*)^p  &=& 
-\frac{1}{n+1} (m^{*})^{n+1} (\overline D \overline D m^*)^{p+1}  + \cdots,
\label{eq:DphicDphicIBP}
\eeq
where the ellipses stand for total superderivatives that vanish in the effective action after integrating
over superspace. 

It is nevertheless useful to re-express ${\cal L}_{\rm eff}$ in a form that does involve $(Dm Dm)(\overline Dm^* \overline Dm^*)$. Making use of the facts that 
$D_\alpha m^* = 0$ and $D^{\dot\alpha} m = 0$ for chiral superfields, and that any product of three chiral or three anti-chiral derivatives vanish,
$D_\alpha m D_\beta m D_\gamma m = 0$ and 
$\overline D_{\dot \alpha} m^* \overline D_{\dot\beta} m^* \overline D_{\dot\gamma} m^* = 0$, 
it follows from eqs.~(\ref{eq:DphiDphiIBP}) and (\ref{eq:DphicDphicIBP}) that the effective auxiliary
field potential contribution to
eq.~(\ref{eq:KJWeffWZ}) can be rewritten in the form
\beq
\int d^2\theta d^2\theta^\dagger J_{\rm eff}
&=&
\int d^2\theta d^2\theta^\dagger  
\frac{1}{16} (Dm Dm) (\overline{D}m^* \overline{D}m^*) 
\,
G_{\rm eff} (m,\, m^*,\, \Delta,\,  \Delta^*)
, 
\phantom{xxx}
\label{eq:KGWeffWZ}
\eeq
in which
\beq
G_{\rm eff} &=& 
\frac{1}{|\Delta|^2}
\frac{\partial^2}{\partial m^* \partial m} J_{\rm eff}
,
\label{eq:GeffJeff}
\eeq
with $\Delta$ and $\Delta^*$ treated as constants in taking the partial derivatives. 
Although $G_{\rm eff}$ and $J_{\rm eff}$ contain equivalent information, $G_{\rm eff}$ is much easier to derive in a direct way from the known $V_{\rm eff}$, and as a consequence it is also easier to express $G_{\rm eff}$ in terms of renormalized master integrals. The complete results for $G_{\rm eff}$ and $J_{\rm eff}$ for the Wess-Zumino model at 1-loop order have been obtained by S.M.~Kuzenko and S.J.~Tyler in refs.~\cite{Kuzenko:2014ypa,Tyler:2013mgu}. In the following I will provide the 2-loop and 3-loop order extensions of those results.

Writing the loop expansion for the effective auxiliary field potential as
\beq
G_{\rm eff} &=& \sum_{\ell=1}^\infty \kappa^\ell G^{(\ell)} ,
\eeq
the scalar effective potential can be expressed, using eqs.~(\ref{eq:Keffloopexp}) and (\ref{eq:KGWeffWZ}) in eq.~(\ref{eq:KJWeffWZ}), as
\beq
V_{\rm eff} &=& |w|^2 - |F + w^*|^2 
%- |\lambda|^2 FF^* \sum_{\ell=1}^\infty \kappa^\ell \frac{\partial^2}{\partial m\partial m^*} K^{(\ell)} 
%\nonumber \\ &&
%- |\lambda|^4 F^2  F^{*2} \sum_{\ell=1}^\infty \kappa^\ell G^{(\ell)}(m, m^*, \lambda F, \lambda F^*)
- |\lambda F|^2 
\sum_{\ell=1}^\infty \kappa^\ell \left (
\frac{\partial^2 K^{(\ell)}}{\partial m^*\partial m}  
+ |\lambda F|^2 \,G^{(\ell)}(m, m^*, \lambda F, \lambda^* F^*) \right )
,
\label{eq:VeffwithGeff}
\eeq
where $F$ is the auxiliary field for $\phi$, and $w = \mu \phi + \frac{1}{2} \lambda \phi^2$ as in eq.~(\ref{eq:superfieldconstraint}),  with $K^{(\ell)}$ known from the previous section,
and $\phi$ is now taken everywhere to be the complex scalar field. Now one eliminates $F$ and $F^*$ through their algebraic equations of motion
$\partial V_{\rm eff}/\partial F = \partial V_{\rm eff}/\partial F^* = 0$, by expanding
\beq
F &=& -w^* + \sum_{\ell=1}^\infty \kappa^\ell F^{(\ell)},
\eeq
and then solve for and eliminating the $F^{(\ell)}$ for $\ell= 1,2,3$ in turn. Then one can solve eq.~(\ref{eq:VeffwithGeff}) for the functions $G^{(\ell)}$ for $\ell= 1,2,3$ in turn, using the known results of $K^{(\ell)}$ from eqs.~(\ref{eq:K1WZ})-(\ref{eq:K3WZ}) above, and the known results for $V_{\rm eff}$ from ref.~\cite{{Martin:2017lqn}}. (Note that the equation determining $G^{(2)}$ depends on the first derivatives
of $G^{(1)}$ with respect to the last two arguments, and similarly the equation determining $G^{(3)}$
depends on the first derivatives of $G^{(2)}$ and the second derivatives of $G^{(1)}$.)
After doing so, the equivalent description of the effective auxiliary field potential
\beq
J_{\rm eff} &=& \sum_{\ell=1}^\infty \kappa^\ell J^{(\ell)} ,
\eeq 
is found by applying eq.~(\ref{eq:GeffJeff}) by integrating $G^{(\ell)}$ twice, with respect to $m$ 
and then $m^*$. The constants of integration are determined by the requirement that $J_{\rm eff}$ is zero
for vanishing $\Delta$ or $\Delta^*$.
Once their functional forms are known, $G_{\rm eff}$ and
$J_{\rm eff}$ can be reinterpreted as superfield quantities.

In order to express the results succinctly, define real superfield combinations
\beq
y &=& x + |\Delta| ,
\qquad\quad
z \>=\> x - |\Delta|  ,
\\
\delta &=& |\Delta|/x .
\eeq
Now one finds, following the procedure described in the previous paragraph at 1-loop order,
\beq
G^{(1)} &=& \frac{1}{|\Delta|^2} \left ( \frac{3}{4} + \frac{1}{2} \lnbar(x) \right )
+ \frac{1}{|\Delta|^4} \left (\frac{1}{2} x^2 \lnbar(x) - \frac{1}{4} y^2 \lnbar(y) - \frac{1}{4} z^2 \lnbar(z) \right )
\\
&=& \frac{1}{4 x^2 \delta^4} \bigl [ 3 \delta^2 - (1 + \delta)^2 \ln(1 + \delta) - (1 - \delta)^2
\ln(1-\delta) \bigr ] ,
\label{eq:G1exact}
\eeq
or, expanding in a series in $\delta$,
\beq
G^{(1)} &=& 
\frac{1}{x^2} \left (
\frac{1}{24} + 
\frac{\delta^2}{120} + 
\frac{\delta^4}{336} +
\frac{\delta^6}{720} +
\frac{\delta^8}{1320} +
\frac{\delta^{10}}{2184} +
\cdots
\right )
. 
\label{eq:G1series}
\eeq 
Applying eq.~(\ref{eq:GeffJeff}) by integrating eq.~(\ref{eq:G1exact}) twice, one obtains the equivalent result
\beq
J^{(1)} &=& \frac{x}{36} \Bigl \{
4 + 3 \delta \, {\rm Li}_2 (\delta) 
- \frac{1}{\delta^2} \left ( 1 + \delta \right )
\left (1 + 7 \delta/2 +  11 \delta^2/2 \right ) \ln (1 + \delta)
\Bigr \} + (\delta \rightarrow -\delta) .
\label{eq:J1exact}
\eeq
Note that $J^{(1)}$ is more complicated than $G^{(1)}$ in the sense that the former necessarily involves dilogarithms, although they contain the same information. Expanding in a series in $\delta$ gives
\beq
J^{(1)} &=& 
x  \left (
\frac{\delta^2}{24}  + 
\frac{\delta^4}{1080} + 
\frac{\delta^6}{8400} +
\frac{\delta^8}{35280} +
\frac{\delta^{10}}{106920} +
\frac{\delta^{12}}{264264} +
\cdots
\right )
. 
\label{eq:J1series}
\eeq 
Equations (\ref{eq:G1exact})-(\ref{eq:J1series})
reproduce the original superfield calculations of refs.~\cite{Tyler:2013mgu,Kuzenko:2014ypa}
for the 1-loop effective auxiliary field potential. (The first term in the expansion of eq.~(\ref{eq:G1series}) had been given earlier in ref.~\cite{Pickering:1996he}.)

At 2-loop order, I find that the procedure outlined above gives
\beq
G^{(2)} &=& 
\frac{\lambda^2 m^{*2} \Delta^* + \lambda^{*2} m^{2} \Delta}{32 |\Delta|^5}
\Bigl (8I(x,x,z) - 8 I(x,x,y) + 3 I(y,y,y) - 3 I(z,z,z) \nonumber \\ &&
+ I(y,z,z) - I(y,y,z) \Bigr )
+  \frac{|\lambda|^2}{16 |\Delta|^4} \Bigl ( 3 x I(y,y,y) + 3 x I(z,z,z) + x I(y,y,z) 
\nonumber \\ &&
+ x I(y,z,z) - 4 z I(x,x,y)
 - 4 y I(x,x,z) - 2 \bigr [ 2 x \lnbar(x) - y \lnbar(y) - z \lnbar(z) \bigl ]^2
\Bigr )
\nonumber \\ && 
+ \frac{|\lambda|^2}{|\Delta|^2} \left (-\frac{1}{4} - \frac{1}{6} c_I + \frac{1}{4} \lnbar^2(x) \right ) ,
\label{eq:G2exact}
\eeq
or, after expanding in a series in $\Delta, \Delta^*$ up to order $|\Delta|^7$,
\beq
G^{(2)} &=& 
\frac{\lambda^2 m^{*2} \Delta^* + \lambda^{*2} m^{2} \Delta}{x^4 \delta^2}
\Biggl [ -\frac{1}{18} c_I 
+ \delta^2 
\left (\frac{2}{27} - \frac{7}{162} c_I  \right )
+ \delta^4 
\left (\frac{11}{72} - \frac{1}{18} c_I \right )
\nonumber \\ &&
\qquad 
+ \delta^6 
\left (\frac{11791}{38880} - \frac{137}{1458} c_I \right )
+ \delta^8 
\left (\frac{68997007}{110224800} - \frac{21661}{118098} c_I \right )
+ \cdots
\Biggr ] 
\nonumber \\ &&
+ 
\frac{|\lambda|^2}{x^2}
\Biggl [\frac{1}{8} 
+ \delta^2  
\left (\frac{539}{2160} - \frac{4}{81} c_I - \frac{1}{120} \lnbar(x) \right )
+ \delta^4
\left (\frac{15313}{34020} - \frac{83}{729} c_I - \frac{1}{168} \lnbar(x) \right )
\nonumber \\ && 
\qquad 
+ \delta^6  
\left (\frac{21558767}{24494400} - \frac{1577}{6561} c_I - \frac{1}{240} \lnbar(x) \right )
+ \cdots \Biggr ]
.
\label{eq:G2series}
\eeq
This expansion can be obtained either using the analytical form of the renormalized master integral 
$I$ for general arguments, which
can be found in eq.~(5.4) of ref.~\cite{Martin:2016bgz} for example, or from its derivatives, which can be found in the ancillary file {\tt derivatives.txt} accompanying the same paper.

The integration of eq.~(\ref{eq:G2exact}) to obtain the exact form of $J^{(2)}$ is much more complicated, and will not be attempted here. Instead, I give the series expanded form through order $|\Delta|^9$, which can be obtained directly from eq.~(\ref{eq:G2series}) by integrating twice according to eq.~(\ref{eq:GeffJeff}):
\beq
J^{(2)} &=& 
\frac{\lambda^2 m^{*2} \Delta^* + \lambda^{*2} m^{2} \Delta}{x}
\Biggl [
\frac{1}{18} c_I + 
\delta^2 
\left (\frac{2}{81} - \frac{7}{486} c_I  \right )
+ \delta^4 
\left (\frac{11}{1080} - \frac{1}{270} c_I \right )
\nonumber \\ &&
\qquad + \delta^6  
\left (\frac{11791}{1360800} - \frac{137}{51030} c_I \right ) 
 + \delta^8 
\left (\frac{68997007}{6944162400} - \frac{21661}{7440174} c_I \right )
+ \cdots
\Biggr ] 
\nonumber \\ && 
+ |\lambda|^2 x 
\Biggl [\frac{1}{8} \delta^2 
+ \delta^4 
\left (\frac{527}{19440} - \frac{4}{729} c_I - \frac{1}{1080} \lnbar(x) \right )
+ \delta^6  
\left (\frac{544}{30375} - \frac{83}{18225} c_I - \frac{1}{4200} \lnbar(x) \right )
\nonumber \\ && 
\qquad + \delta^8 
\left (\frac{21529607}{1200225600} - \frac{1577}{321489} c_I - \frac{1}{11760} \lnbar(x) \right )
+ \cdots
\Biggr ].
\label{eq:J2series}
\eeq
Note that, unlike the 1-loop contribution,  
the 2-loop effective auxiliary field potential contains odd powers of $\Delta$, $\Delta^*$.
In the $J^{(2)}$ version, the expansion starts from terms linear in $\Delta$ or $\Delta^*$, and each
term contains at most one more power of $\Delta$ than $\Delta^*$ or vice versa. The terms with
odd powers do not contain logarithms of the renormalization scale.

The 3-loop effective auxiliary field potential is much more complicated, but I have obtained it in the form
\beq
G^{(3)} &=& 
\frac{|\lambda|^4 x^2}{|\Delta|^4} G^{(3)}_a
+ \frac{|\lambda|^2 \bigl ( \lambda^2 m^{*2} \Delta^* + \lambda^{*2} m^{2} \Delta \bigr ) x}{|\Delta|^5} G^{(3)}_b
+ \frac{\lambda^4 m^{*4} \Delta^{*2} + \lambda^{*4} m^{4} \Delta^2}{|\Delta|^6} G^{(3)}_c
,
\phantom{xxx}
\label{eq:G3genform}
\eeq
where the dimensionless functions $G^{(3)}_{a,b,c}$ depend only on $x$, $\delta$, and $Q$. They are combinations of the renormalized master integrals $\lnbar$, $I$, $\overline I$, $F$, $G$, $H$, and $K$ defined in ref.~\cite{Martin:2016bgz}, which also provides for their numerical evaluation through known differential equations. The squared-mass arguments of the master integral functions are always $x$, $y$, or $z$, in a way analogous to eq.~(\ref{eq:G2exact}). The explicit results for $G^{(3)}_{a,b,c}$ are too long to show in print here, and so are given in the ancillary file {\tt G3abc.txt} distributed with this paper in a form suitable for use in computer programs.

The expansion of eq.~(\ref{eq:G3genform}) in a series in $|\Delta|$ is made slightly trickier by the fact
that in the general expressions for derivatives of the function $F(w,x,y,z)$ found in the ancillary file {\tt derivatives.txt} of ref.~\cite{Martin:2016bgz}, the numerators and denominators both
vanish for $w=x=y=z$. To get around this, first one needs the special values
\beq
F(x,x,x,x) &=& x \left [\frac{53}{12} + \frac{13}{4} \lnbar(x) - 4 \lnbar^2(x) + \lnbar^3(x) \right ],
\\
\left [\frac{\partial}{\partial y} F(y,x,x,x)\right ] \Bigl |_{y=x} &=& \frac{26}{3} - 7 \zeta_3 - \frac{31}{4} \lnbar(x) + 2 \lnbar^2(x),
\\
\left [\frac{\partial}{\partial y} F(x,x,x,y)\right ] \Bigl |_{y=x} &=& 
-\frac{1}{3} + \frac{7}{3} \zeta_3 + \lnbar(x) - \lnbar^2(x) + \frac{1}{3} \lnbar^3(x) ,
\eeq
which can all be obtained using the analytical formula for $F(x,x,y,y)$ in eq.~(5.66) of ref.~\cite{Martin:2016bgz}. Then arbitrary higher derivatives evaluated for equal squared mass arguments can be obtained by taking limits of the derivatives evaluated for two different squared mass arguments.

I find for the expansion of eq.~(\ref{eq:G3genform}), through order $|\Delta|^4$,
\beq
G^{(3)} &=& \frac{|\lambda|^4}{x^2} 
\Biggl \{
-\frac{1}{4} + \frac{2}{9} c_I - \frac{463}{864} \zeta_3 + \frac{3}{16} \lnbar(x) 
+ \delta^2 
\biggl (-\frac{83}{240} - \frac{21767}{6480} c_I + \frac{50491}{4860} \zeta_3 
\nonumber \\ && \qquad
+ \left[ \frac{539}{4320} - \frac{2}{81} c_I \right ] \lnbar(x) - \frac{1}{480} \lnbar^2(x)
\biggr ) 
+ \delta^4 \biggl (
-\frac{62933039}{17418240} - \frac{981817}{25515} c_I 
\nonumber \\ && \qquad
+ \frac{7273176425}{62705664} \zeta_3
+ \left [-\frac{15313}{68040} + \frac{83}{1458} c_I \right ] \lnbar(x)
+ \frac{1}{672} \lnbar^2(x)
\biggr )
+ \cdots
\Biggr \}
\nonumber \\ &&
+ 
\frac{|\lambda|^2\bigl ( \lambda^2 m^{*2} \Delta^* + \lambda^{*2} m^{2} \Delta ) }{x^4 \delta^2} 
\Biggl \{
-\frac{1}{9} c_I + \frac{1}{2} \zeta_3 - \frac{1}{9} c_I \lnbar(x)
+\delta^2  
\biggl (\frac{1}{36} - \frac{5}{6} c_I + \frac{1457}{648} \zeta_3 
\nonumber \\ && \qquad
+ \left[ \frac{2}{27} - \frac{7}{162} c_I \right ] \lnbar(x) 
\biggr )
+\delta^4  
\biggl (
 -\frac{16199}{25920} - \frac{19727}{2430} c_I + \frac{744953}{31104} \zeta_3
\biggr )
 + \cdots
\Biggr \}
\nonumber \\ &&
+ 
\frac{\lambda^4 m^{*4} \Delta^{*2} + \lambda^{*4} m^{4} \Delta^2}{x^6 \delta^2} 
\Biggl \{
-\frac{1}{6} c_I + \frac{91}{144} \zeta_3 
+ \delta^2 
\biggl (-\frac{7}{72} - \frac{55}{81} c_I + \frac{8663}{3888} \zeta_3 \biggr )
\nonumber \\ && \qquad
+ \delta^4 
\biggl (-\frac{711425}{995328} - \frac{4279}{972} c_I + \frac{244176635}{17915904} \zeta_3 
\biggr )
 + \cdots
\Biggr \}
.
\phantom{xxx}
\label{eq:G3series}
\eeq
Applying eq.~(\ref{eq:GeffJeff}), the first terms in the series expansion of the 3-loop contribution to the alternative version 
$J_{\rm eff}$ are, through order $|\Delta|^6$,
\beq
J^{(3)} &=& |\lambda|^4 x
\Biggl \{
\delta^2 \biggl (\frac{1}{8} + \frac{2}{9} c_I - \frac{463}{864} \zeta_3 + \frac{3}{16} \lnbar(x) \biggr )
+ \delta^4  
\biggl (-\frac{1711}{58320} - \frac{65621}{174960} c_I + \frac{50491}{43740} \zeta_3 
\nonumber \\ && \qquad
+ \left[ \frac{527}{38880} - \frac{2}{729} c_I \right ] \lnbar(x) - \frac{1}{4320} \lnbar^2(x)
\biggr ) 
+ \delta^6 \biggl (
-\frac{35830483}{241920000} - \frac{981236}{637875} c_I 
\nonumber \\ && \qquad
+ \frac{290927057}{62705664} \zeta_3
+ \left [-\frac{272}{30375} + \frac{83}{36450} c_I \right ] \lnbar(x)
+ \frac{1}{16800} \lnbar^2(x)
\biggr )
+ \cdots
\Biggr \}
\nonumber \\ &&
+ 
\frac{|\lambda|^2\bigl ( \lambda^2 m^{*2} \Delta^* + \lambda^{*2} m^{2} \Delta ) }{x} 
\Biggl \{
\frac{1}{9} c_I - \frac{1}{2} \zeta_3 + \frac{1}{9} c_I \lnbar(x)
+\delta^2  
\biggl (\frac{41}{972} - \frac{433}{1458} c_I + \frac{1457}{1944} \zeta_3 
\nonumber \\ && \qquad
+ \left[ \frac{2}{81} - \frac{7}{486} c_I \right ] \lnbar(x) 
\biggr )
+\delta^4  
\biggl (
 -\frac{16199}{388800} - \frac{19727}{36450} c_I + \frac{744953}{466560} \zeta_3
\biggr )
 + \cdots
\Biggr \}
\nonumber \\ &&
+ 
\frac{\lambda^4 m^{*4} \Delta^{*2} + \lambda^{*4} m^{4} \Delta^2}{x^3} 
\Biggl \{
\frac{1}{18} c_I - \frac{91}{432} \zeta_3 
+ \delta^2 
\biggl (-\frac{7}{360} - \frac{11}{81} c_I + \frac{8663}{19440} \zeta_3 
\biggr )
\nonumber \\ && \qquad
+ \delta^4 
\biggl (-\frac{711425}{20901888} - \frac{4279}{20412} c_I + \frac{244176635}{376233984} \zeta_3 
\biggr )
 + \cdots
\Biggr \}
.
\phantom{xxx}
\label{eq:J3series}
\eeq
Note that the expansion for $J^{(3)}$ again starts from terms
linear in $\Delta$ or $\Delta^*$. However, unlike at 2-loop order, $G^{(3)}$ or $J^{(3)}$ contain terms with up to two more powers of $\Delta$ than $\Delta^*$.

\section{Renormalization group invariance\label{sec:rge}}
\setcounter{equation}{0}
\setcounter{figure}{0}
\setcounter{table}{0} 
\setcounter{footnote}{1}

The effective action must be invariant with respect to changes in the arbitrary \MSbar renormalization scale $Q$. This implies that
\beq
Q \frac{d}{dQ} \left [\int d^2\theta d^2\theta^\dagger \, L_{\rm eff} + \left (\int d^2 \theta \, W  + {\rm c.c.}\right ) \right ] = 0
,
\eeq
where the differentiation includes both the dependence on $Q$ of the running \MSbar background fields and coupling parameters and the explicit (though not listed among the function arguments) dependence on $Q$ contained in the loop integral functions $\lnbar(x)$, $I(x,y,z)$, etc. It is useful and
instructive to check this renormalization group invariance, which in superspace is realized
separately for each of the superpotential, the effective K\"ahler potential, and the effective auxiliary potential.
As in the rest of this paper, I will exclude the parts of the superspace effective action that contain explicit spacetime derivatives acting on superfields.

The non-renormalization theorem tells us that the perturbative effective superpotential
is simply the tree-level one, and so
has no explicit $Q$ dependence from loop corrections. Therefore, for the superpotential $W(\phi_i)$ with the form in eq.~(\ref{eq:WXi}), it must be true that
 \beq
Q \frac{dW}{dQ}  \>=\> \sum_X \beta_X \frac{\partial W}{\partial X}  &=& 0,
\label{eq:QdQWeq0}
\eeq
where $X$ runs over the \MSbar background fields and coupling parameters,
$X = \phi_i,\> \mu^{ij},\> \lambda^{ijk}$,
and the corresponding beta functions are 
\beq
\beta_X &=& Q\frac{dX}{dQ} \>=\> \sum_{\ell=1}^\infty \kappa^\ell \beta_X^{(\ell)}.
\eeq
The vanishing of eq.~(\ref{eq:QdQWeq0}) is realized as 
\beq
\beta_{\phi_i} &=& -\gamma_i^n \phi_n,
\label{eq:betaphii}
\\[4pt]
\beta_{\mu^{ij}} &=& \gamma_n^i \mu^{nj}  + \gamma_n^j \mu^{in} ,
\label{eq:betamuij}
\\[4pt]
\beta_{\lambda^{ijk}} &=& \gamma_n^i \lambda^{njk}  + \gamma_n^j \lambda^{ink}  +  \gamma_n^k \lambda^{ijn},
\label{eq:betalambdaijk}
\eeq
where the chiral superfield anomalous dimension can be expanded as
\beq
\gamma_{i}^j &=& \sum_{\ell = 1}^\infty \kappa^\ell \gamma_{i}^{(\ell) j},
\\
\gamma_{i}^{(1) j} &=& \frac{1}{2} \lambda_{ikl}\lambda^{jkl},
\label{eqgammaij1}
\\
\gamma_{i}^{(2) j} &=& -\frac{1}{2} \lambda_{ikl}\lambda^{jkn}\lambda^{lpq}\lambda_{npq},
\label{eqgammaij2}
\\
\gamma_{i}^{(3) j} &=& 
-\frac{1}{8} \lambda_{ikl} \lambda^{jpq} \lambda^{kmn} \lambda_{pmn} \lambda^{lrs} \lambda_{qrs}
-\frac{1}{4} \lambda_{ikl} \lambda^{jkm} \lambda^{lnp} \lambda_{snp} \lambda^{sqr} \lambda_{mqr} 
\nonumber \\
&&
+ \lambda_{ikl} \lambda^{jkm} \lambda^{lnp} \lambda_{mnq} \lambda^{qrs} \lambda_{prs}
+ \frac{3}{2} \zeta_3 \lambda_{ikl} \lambda^{jpq} \lambda^{kmn} \lambda^{lrs} \lambda_{pmr} \lambda_{qns}
,
\label{eqgammaij3}
\eeq
at 3-loop order, as found in refs.~\cite{West:1984dg,Avdeev:1982jx,Jack:1996qq}.

For the K\"ahler potential, the renormalization group invariance condition becomes
\beq
Q\frac{\partial K^{(\ell)}}{\partial Q}  \,+\, \sum_{n=0}^{\ell - 1} \left ( \sum_X \beta_X^{(\ell-n)} \frac{\partial K^{(n)}}{\partial X}  \right ) 
&=& 0
\label{eq:RGcheckKell}
\eeq
for each loop order $\ell = 1,2,3$, and now
\beq
X = \phi_i,\> 
\phi^{*i},\> 
\mu^{ij},\> 
\mu_{ij},\> 
\lambda^{ijk},\> 
\lambda_{ijk}
,
\eeq
and it is important that an integration $\int d^2\theta d^2\theta^\dagger$ is understood. The first term in eq.~(\ref{eq:RGcheckKell}) reflects the explicit $Q$ dependence of the K\"ahler potential coming
from the loop integral functions. To evaluate it, and similar terms below, one needs only the identities
\beq
Q \frac{\partial}{\partial Q} \lnbar(x) &=& -2,
\label{eq:QdQlnbar}
\\
Q \frac{\partial}{\partial Q} I(x,y,z) &=& 2 x \lnbar(x) + 2 y \lnbar(y) + 2 z \lnbar(z) - 4 x - 4 y - 4 z,
\label{eq:QdQI}
\\
Q \frac{\partial}{\partial Q} \overline{I}(x,y,z) &=& 2 - 2 \lnbar(x),
\label{eq:QdQIbar}
\\
Q \frac{\partial}{\partial Q} F(w,x,y,z) &=& 2 \lnbar(w) \left [
2x + 2 y + 2 z  -w
-x \lnbar(x) - y \lnbar(y) - z \lnbar(z)\right ]
\nonumber \\ && 
+ 2 x \lnbar(x) + 2 y \lnbar(y) + 2 z \lnbar(z) -4x - 4 y - 4 z + 11 w/2
,
\label{eq:QdQF}
\\
Q \frac{\partial}{\partial Q} G(v,w,x,y,z) &=& 2 I(v,w,x) + 2 I(v,y,z)  + 2 w \lnbar(w) + 2 x \lnbar(x) + 2 y \lnbar(y) 
\nonumber \\ && 
+ 2 z \lnbar (z) + 2 v - 6 w - 6 x - 6 y - 6 z,
\label{eq:QdQG}
\\
Q \frac{\partial}{\partial Q} H(u,v,w,x,y,z) &=& 12 \zeta_3,
\label{eq:QdQH}
\\
Q \frac{\partial}{\partial Q} K(u,v,w,x,y,z) &=& 2 \overline{I}(u,v,w,x) + 2 \overline{I}(u,v,y,z) - 2,
\label{eq:QdQK}
\eeq
as found in ref.~\cite{Martin:2016bgz}.

At 1-loop order, we have from eqs.~(\ref{eq:K0gen}), (\ref{eq:betaphii}), and (\ref{eqgammaij1}) that
\beq
\sum_X \beta_X^{(1)} \frac{\partial K^{(0)}}{\partial X}  = -\phi^{*i} \lambda_{ikl} \lambda^{klj} \phi_j, 
\label{eq:QdQK0gen1}
\eeq
while eqs.~(\ref{eq:K1gen}) and (\ref{eq:QdQlnbar}) give
\beq
Q\frac{\partial K^{(1)}}{\partial Q}  &=& \sum_k x_k
.
\label{eq:QdQK1gen1}
\eeq
Equations (\ref{eq:QdQK0gen1}) and (\ref{eq:QdQK1gen1}) cancel after integrating over superspace, since
\beq
\int d^2\theta d^2\theta^\dagger \,\phi^{*i} \lambda_{ikl} \lambda^{klj} \phi_j
&=& 
\int d^2\theta d^2\theta^\dagger \,
(\mu_{kl} + \phi^{*i} \lambda_{ikl})(\mu^{kl} +  \lambda^{klj}\phi_{j})
\>=\>
\int d^2\theta d^2\theta^\dagger \, m_{kl} m^{lk}
\nonumber \\ &=&
\int d^2\theta d^2\theta^\dagger \sum_k x_k,
\phantom{xxx}
\eeq
where the first equality follows because the terms involving $\mu_{kl}$ and $\mu^{kl}$ do not contribute. This verifies eq.~(\ref{eq:RGcheckKell}) with $\ell=1$.

At 2-loop order, eqs.~(\ref{eq:K0gen}), (\ref{eq:betaphii}), and (\ref{eqgammaij2}) yield
\beq
\sum_X \beta_X^{(2)} \frac{\partial K^{(0)}}{\partial X}  \,=\, 
\phi^{*i} \lambda_{ikl} \lambda^{jkn} \lambda^{lpq} \lambda_{npq} \phi_j
\,=\, m_{kl} m^{kn} \lambda^{lpq} \lambda_{npq}
\,=\, x_n Y^{npq} Y_{npq}
, 
\label{eq:QdQK0gen2}
\eeq
where the second equality again relies on the implicit integration over superspace, and the last equality
follows from appropriate insertions of $U$ and $U^\dagger$ as defined by eq.~(\ref{eq:defineUA})-(\ref{eq:defineUY}).
Also, from eqs.~(\ref{eq:K1gen}), (\ref{eq:betaphii})-(\ref{eq:betalambdaijk}), and (\ref{eqgammaij1}), one finds
\beq
\sum_X \beta_X^{(1)} \frac{\partial K^{(1)} }{\partial X} \,=\, 
x_i \left [1 - \lnbar(x_i) \right ] Y^{ijk} Y_{ijk} 
.
\eeq
Finally, eqs.~(\ref{eq:K2gen}) and (\ref{eq:QdQI}) yield
\beq
Q\frac{\partial K^{(2)}}{\partial Q} &=& x_i \left [\lnbar(x_i) - 2 \right ] Y^{ijk} Y_{ijk} .
\label{eq:QdQK2gen2}
\eeq
The sum of eqs.~(\ref{eq:QdQK0gen2})-(\ref{eq:QdQK2gen2}) vanishes as required by eq.~(\ref{eq:RGcheckKell}) with $\ell=2$.

Moving on to the 3-loop order check, eqs.~(\ref{eq:K0gen}), (\ref{eq:betaphii}), and (\ref{eqgammaij3}) yield,
after once again using the implicit integration over superspace and inserting $U$ and $U^\dagger$ appropriately,
\beq
\sum_X \beta_X^{(3)} \frac{\partial K^{(0)}}{\partial X} &=& 
\frac{1}{4} Y^{ikl} Y_{jkl} Y^{i'np} Y_{j'np} M_{ii'} M^{jj'}
+ Y^{ikl} Y_{jkl} Y_{inp} Y^{jnp}\left (\frac{1}{2} x_i - 2 x_k \right )
\nonumber \\ &&
- 3 \zeta_3 Y^{ijl} Y_{jkn} Y^{i'kp} Y_{ln'p} M_{ii'} M^{nn'},
\label{eq:QdQK0gen3}
\eeq
while eqs.~(\ref{eq:K1gen}), (\ref{eq:betaphii})-(\ref{eq:betalambdaijk}), and (\ref{eqgammaij2}) give
\beq
\sum_X \beta_X^{(2)} \frac{\partial K^{(1)}}{\partial X} &=& x_i \left [\lnbar(x_i) - 1 \right ] Y^{ijk} Y_{ijl} Y^{lnp} Y_{knp} 
,
\eeq
and eqs.~(\ref{eq:K2gen}), (\ref{eq:betaphii})-(\ref{eq:betalambdaijk}), and (\ref{eqgammaij1}) result in
\beq
\sum_X \beta_X^{(1)} \frac{\partial K^{(2)}}{\partial X} &=& 
- \frac{1}{4} \overline{I} (x_i, x_j, x_k, x_l) Y^{ikl} Y_{jkl}
\left [ (x_i + x_j)  Y_{inp} Y^{jnp}
+ 2 M_{ii'} M^{jj'} Y^{i'np} Y_{j'np}
\right ]
\nonumber \\
&&
+ \frac{1}{4} I(x_i, x_j, x_k) \left [
Y^{ijk} Y_{ijl} Y^{lnp} Y_{knp} + Y_{ijk} Y^{ijl} Y_{lnp} Y^{knp} \right ]
.
\label{eq:QdQK2gen3}
\eeq
As required by eq.~(\ref{eq:RGcheckKell}) with $\ell=3$,
the sum of eqs.~(\ref{eq:QdQK0gen3})-(\ref{eq:QdQK2gen3}) cancels against  $Q \frac{\partial}{\partial Q} K^{(3)}$, which is
obtained immediately by applying eqs.~(\ref{eq:QdQG})-(\ref{eq:QdQK}) to eq.~(\ref{eq:K3gen}).

In a similar way, one can check the renormalization group invariance of the effective auxiliary field potential for the Wess-Zumino model, in both the $J_{\rm eff}$ and $G_{\rm eff}$ incarnations. The necessary conditions can be written as
\beq
0 &=& Q\frac{\partial J^{(\ell)}}{\partial Q} \,+\, \sum_{n=1}^{\ell-1} \sum_X \beta_X^{(\ell-n)} 
\frac{\partial J^{(n)}}{\partial X},
\label{eq:QdQJell}
\\
0 &=& Q\frac{\partial G^{(\ell)}}{\partial Q} \,+\, \sum_{n=1}^{\ell-1} \left [ 8 \gamma^{(\ell - n)} + \sum_X \beta_X^{(\ell-n)} \frac{\partial}{\partial X} \right ] G^{(n)},
\label{eq:QdQGell}
\eeq
where the \MSbar running parameters are $X = \phi, \phi^*, m, m^*, \lambda, \lambda^*, x, \Delta, \Delta^*, \delta$. The beta functions for them are  
\beq
-\frac{\beta^{(\ell)}_\phi}{\phi} \,=\, 
\frac{\beta^{(\ell)}_m}{2 m} \,=\,
\frac{\beta^{(\ell)}_\lambda}{3 \lambda} \,=\,
\frac{\beta^{(\ell)}_x}{4 x} \,=\,
\frac{\beta^{(\ell)}_\Delta}{2 \Delta} \,=\,
-\frac{\beta^{(\ell)}_\delta}{2 \delta} \,=\,
\gamma^{(\ell)},
\eeq
where the 3-loop contributions to the superfield anomalous dimension are
\beq
\gamma^{(1)} &=& \frac{1}{2} |\lambda|^2,
\\
\gamma^{(2)} &=& -\frac{1}{2} |\lambda|^4,
\\
\gamma^{(3)} &=& \left ( \frac{3}{2} \zeta_3 + \frac{5}{8} \right ) |\lambda|^6 .
\eeq
Note that the sums in eqs.~(\ref{eq:QdQJell}) and (\ref{eq:QdQGell}) start from $n=1$, because by assumption there is no tree-level part of $J_{\rm eff}$ or $G_{\rm eff}$. This explains why the expressions for
$G^{(1)}$ and $J^{(1)}$ in eqs.~(\ref{eq:G1exact})-(\ref{eq:J1series}) have no explicit dependence on $Q$,
since the $\ell=1$ cases for eqs.~(\ref{eq:QdQJell}) and (\ref{eq:QdQGell}) are trivial except for the first term. It is more complicated but straightforward to check eq.~(\ref{eq:QdQGell}) in the $\ell=2$ and $\ell=3$ cases as well, using the exact forms for 
$G^{(1)}$ and $G^{(2)}$ in eqs.~(\ref{eq:G1exact}) and (\ref{eq:G2exact}) and $G^{(3)}$ in the ancillary file {\tt G3abc.txt}, together with the identities in eqs.~(\ref{eq:QdQlnbar})-(\ref{eq:QdQK}) above, and derivative formulas in the ancillary file {\tt derivatives.txt} of ref.~\cite{Martin:2016bgz}.
The check is even easier to do for the series expanded forms in eqs.~(\ref{eq:G1series}), (\ref{eq:G2series}) and (\ref{eq:G3series}), where all explicit $Q$ dependence is due to the $\lnbar(x)$ function. Similarly,
it is easy to check eq.~(\ref{eq:QdQJell}) using the series
expanded forms for $J_{\rm eff}$ found in eqs.~(\ref{eq:J1series}), (\ref{eq:J2series}), and (\ref{eq:J3series}).

\vspace{0.3cm}

\section{Outlook\label{sec:outlook}}
\setcounter{equation}{0}
\setcounter{figure}{0}
\setcounter{table}{0} 
\setcounter{footnote}{1}

\vspace{0.3cm}

I have obtained the three-loop contribution to the K\"ahler effective potential for
a renormalizable theory with only chiral supermultiplets, in eq.~(\ref{eq:K3gen}). The simplicity of the result highlights the usefulness of renormalized $\epsilon$-finite master loop integrals, which contain the subtractions of all sub-divergences in a particular way (see \cite{Martin:2021pnd} for a more general discussion). The renormalized integral $I(x,y,z)$ introduced first in 
eq.~(4.27) of ref.~\cite{Ford:1992pn} (where it was called $\hat I(x,y,z)$) allows the 2-loop K\"ahler potential to be written as a single term, and similarly the 3-loop contribution can be written in terms of just four terms involving the three renormalized integrals $G(v,w,x,y,z)$, $H(u,v,w,x,y,z)$, and $K(u,v,w,x,y,z)$. 
However, the simplicity of the result for the 3-loop K\"ahler potential seems to require further explanation, as eq.~(\ref{eq:K3gen}) does not include all terms that one could have imagined including. It might be useful and interesting to understand this by revisiting the calculation directly in terms of superfields, rather than the indirect method of inference used here based on expansion of the scalar component potential to quadratic order in the auxiliary fields.

In order to obtain the complete scalar effective potential from superspace, it is also necessary to have
the effective auxiliary field potential, which contains superderivatives but not spacetime derivatives. This was obtained at 2-loop and 3-loop orders for the special case of the Wess-Zumino model for a single chiral superfield in eqs.~(\ref{eq:G2exact}) and (\ref{eq:G3genform}) and the ancillary electronic file {\tt G3abc.txt}. Here the results are again most neatly written in terms of renormalized $\epsilon$-finite master integrals, but are significantly more complicated than for the K\"ahler potential, with a structure comparable to the full scalar effective potential from which they were inferred. The general structure is perhaps of greater interest than the numerical coefficients. In this regard, the 2-loop and 3-loop order contributions have terms linear in $\Delta$ and $\Delta^*$ which are absent in the 1-loop part as found originally in refs.~\cite{Kuzenko:2014ypa,Tyler:2013mgu}. More generally, it seems that at $\ell$-loop order, the effective auxiliary field potential
contains terms with up to $(\ell-1)$ more powers of $\Delta$ than $\Delta^*$ and vice versa. Again, it would be interesting to elucidate these properties using a more direct superspace calculation.

The same method used in this paper should work at any loop order, in principle. However, the most important practical obstacle is that the 4-loop scalar effective potential (which would be necessary as an input for the
calculation) is not yet known, even for the Wess-Zumino model with
only one chiral superfield.

In this paper, I have not attempted to find comparable results for supersymmetric gauge theories, which of
course are highly relevant for attempts to explain the real world, as well as applications to extended supersymmetric models. The method used here runs into the problem that existing results for the scalar effective potential are based on the non-supersymmetric Landau gauge fixing, while the manifestly supersymmetric effective K\"ahler potential should make use of some supersymmetric gauge-fixing procedure. It would be interesting to generalize the 3-loop results here to the case of supersymmetric gauge theories by conquering these gauge-fixing issues. Once again a direct superspace calculation seems in order.

\vspace{0.3cm}

Acknowledgments: I thank an anonymous referee for pointing out a typo in eq.~(\ref{eq:M2s}), and several useful suggestions. This work is supported by the National Science Foundation grant with award number 2310533.

%%%%%%%%%%%%%%%%%%%%%%%%%%%%%%%%%%%%%%%%%%%%%%%%%%%%%%%%%%%%%%%%%%%%

\end{document}